\documentclass[11pt]{article}
\usepackage[utf8]{inputenc}
\usepackage[T1]{fontenc}
\usepackage{graphicx}
\usepackage{longtable}
\usepackage{wrapfig}
\usepackage{rotating}
\usepackage[normalem]{ulem}
\usepackage{amsmath}
\usepackage{amssymb}
\usepackage{capt-of}
\usepackage{hyperref}
\usepackage[citestyle=authoryear-icomp,bibstyle=authoryear, hyperref=true,backref=true,maxcitenames=3,url=true,backend=bibtex]{biblatex}
\usepackage{multirow}
\usepackage{amsmath}
\usepackage{morefloats}
\usepackage{setspace}
\usepackage{graphicx}
\usepackage{float}
\usepackage{changepage}
\author{Matt Brigida\thanks{Chief Economist, Algorand Foundation \& SUNY Polytechnic Institute.  The insights, contributions and comments of Simon Bonanno, Marc Llopart-Enajas and Michele Treccani among others are greatly acknowledged.}}
\date{\today}
\title{The Surprising Irrelevance of Total-Value-Locked on Cryptocurrency Returns}
\hypersetup{
 pdfauthor={Matt Brigida\thanks{Chief Economist, Algorand Foundation (matthew.brigida@algorand.foundation) \& SUNY Polytechnic Institute (matthew.brigida@sunypoly.edu).  The insights, contributions and comments of Simon Bonanno, Marc Llopart-Enajas and Michele Treccani among others are greatly acknowledged.}},
 pdftitle={The Surprising Irrelevance of Total-Value-Locked on Cryptocurrency Returns},
 pdfkeywords={},
 pdfsubject={},
 pdfcreator={Emacs 30.1 (Org mode 9.7.19)}, 
 pdflang={English}}
\begin{document}

\maketitle
\begin{abstract}
A common assumption in cryptocurrency markets is a positive relationship between total-value-locked (TVL) and cryptocurrency returns.  To test this hypothesis we examine whether the returns of TVL-sorted portfolios can be explained by common cryptocurrency factors. We find evidence that portfolios formed on TVL exhibit returns that are linear functions of aggregate crypto market returns, that is they can be replicated with appropriate weights on the crypto market portfolio.  Thus, strategies based on TVL can be priced with standard asset pricing tools.  This result holds true both for total TVL and a simple TVL measure that removes a number of ways TVL may be overstated.
\end{abstract}
\vspace*{1cm}



\clearpage
\section{Introduction}
\label{sec:org6e161cc}

Decentralized finance (DeFi) is a set of blockchain-based financial services which do not require financial intermediaries. Users can lend, trade, and invest directly with each other using various protocols. Total-value-locked (TVL) is a measure of the total value of cryptocurrency assets deposited in a particular DeFi ecosystem's smart contracts. Thus, TVL measures investment and is commonly viewed as a sign of user confidence in a DeFi platform. A higher TVL means user are willing to lock more of their assets in a protocol, which is consistent with greater trust. This trust can translate into increased platform usage and potentially influence the returns of cryptocurrencies associated with that platform.  Total TVL values have increased from less than a billion in early 2020, to over \$116 billion\footnote{From DefiLlama.com, and excluding doublecounting.} as of May 2025.   

In this analysis we construct various portfolios according to cryptocurrency TVL levels, and test whether these portfolio returns are explainable by crypto factor models.  That is we test whether the space of TVL portfolios can be spanned by linear combinations of a small set of crypto factors.  If they can be spanned, then they can be replicated using appropriate weights on the crypto factors. The idea is that if TVL levels affect a cryptocurrency's returns, then we should be able to form portfolios based on TVL which earn a positive alpha when controlling for standard crypto factors.  We use both total TVL, and TVL adjusted for common transactions which overstate TVL.  We include both because, while arguably the adjusted value is more accurate, total TVL is the more widely cited.

Our analysis is similar to \cite{liu2022common} who uses the same \cite{FAMA_1996} empirical method to find nine cryptocurrency factors can be use to form long-short crypto portfolios with significant returns.  They then show these portfolio returns are explained by at most a three factor crypto model with no significant alpha.

\cite{luo2024piercing} hypothesize that the doublecounting in TVL may increase financial contagion risk between cryptocurrencies.  If this is true, we may be able to construct portfolios based on TVL that generate alpha.  Conversely, if market participants discount how TVL reacts to downturns, then TVL-sorted portfolios should be linear functions of common crypto factors.

\cite{saggese2025towards} attempted to verify TVL values for 939 DeFi projects deployed on Ethereum. To do so they created a new TVL measure which they called verifiable Total Value Locked (vTVL).  They conclude there is substantial difference between their vTVL and widely published TVL values.  This may be a factor causing TVL to be ignored by market participants when determining crypto prices. 

Using a GARCH model on daily TVL to market cap and price data \cite{pantelidis2024evaluating} find evidence of a negative relationship between a currency's TVL and returns. Alternatively, \cite{grande2025trust} define TVL to market capitalization bands, and find that if a cryptocurrency's TVL to market capitalization exceeds the upper band threshold, this predicts a positive change in the cryptocurrency's price. They find crossing the lower band may predict lower prices.  Within the band there is no relationship.  While these analyses attempt to find significant relationships between TVL and returns, we specifically test whether portfolio strategies based on TVL generate anomalous returns in the cryptocurrency market.  Moreover our approach reduces idiosyncratic risk in our regressions by using portfolio rather than individual crypto returns.  Additionally forming portfolios rather than relying on single cryptocurrency returns lessens measurement error from microstructure issues and outliers.

The remainder of the paper is organized as follows.  Section 2 describes our cryptocurrency data, crypto market returns data, and various portfolios formed on TVL, crypto size, and momentum.  Section 3 provides factor regression results across TVL types for all cryptocurrencies and also for Level 1 token only.  Section 4 discusses implications and concludes.
\section{Data}
\label{sec:org79cb8ed}

Our cryptocurrency data set is comprised of weekly returns over a sample period from January 2, 2023 through December 31, 2024.  To ensure our estimates are not affected by a survivorship bias, we gather returns for all cryptocurrencies that were in the top 100 by market cap at any point in our sample period.  This set of cryptocurrencies that have been in the top 100 make up over 97\% of the market capitalization of the top 1000 cryptocurrencies.  So by increasing the sample size we gain very little in market capitalization, however our sample becomes much more exposed to microstructure and price discovery issues.  Note, we exclude Bitcoin, because TVL is not meaningful and its market capitalization dwarfs most other cryptocurrencies.  We also exclude stablecoins.  Using this method our sample contains 335 unique cryptocurrencies.  Cryptocurrency prices were gathered via the Coinmarketcap Application Programming Interface.

Figures \ref{fig:org9f9836c} and \ref{fig:org74a7ad0} below show TVL to market capitalization for both our entire sample of cryptocurrencies and for the Level 1 subset over our sample period. On average the total TVL measure to market cap declined from about 10.5\% to 9\% over our sample.  Over the same period the simple measure of TVL (Total TVL ex values from staking, pool2, governance tokens, borrows, double count, liquid staking, and vesting) declined from 6\% to a little over 4\%.

\begin{figure}[htbp]
\centering
\includegraphics[width=.9\linewidth]{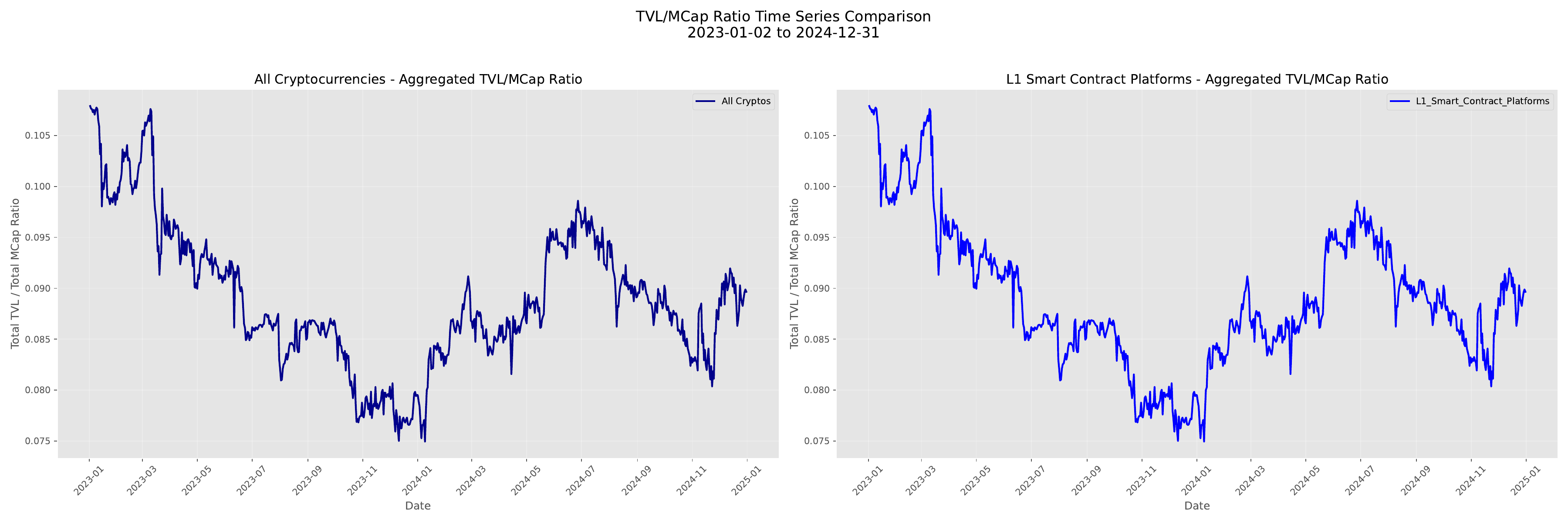}
\caption{\label{fig:org9f9836c}Total TVL to Market Capitalization}
\end{figure}

\begin{figure}[htbp]
\centering
\includegraphics[width=.9\linewidth]{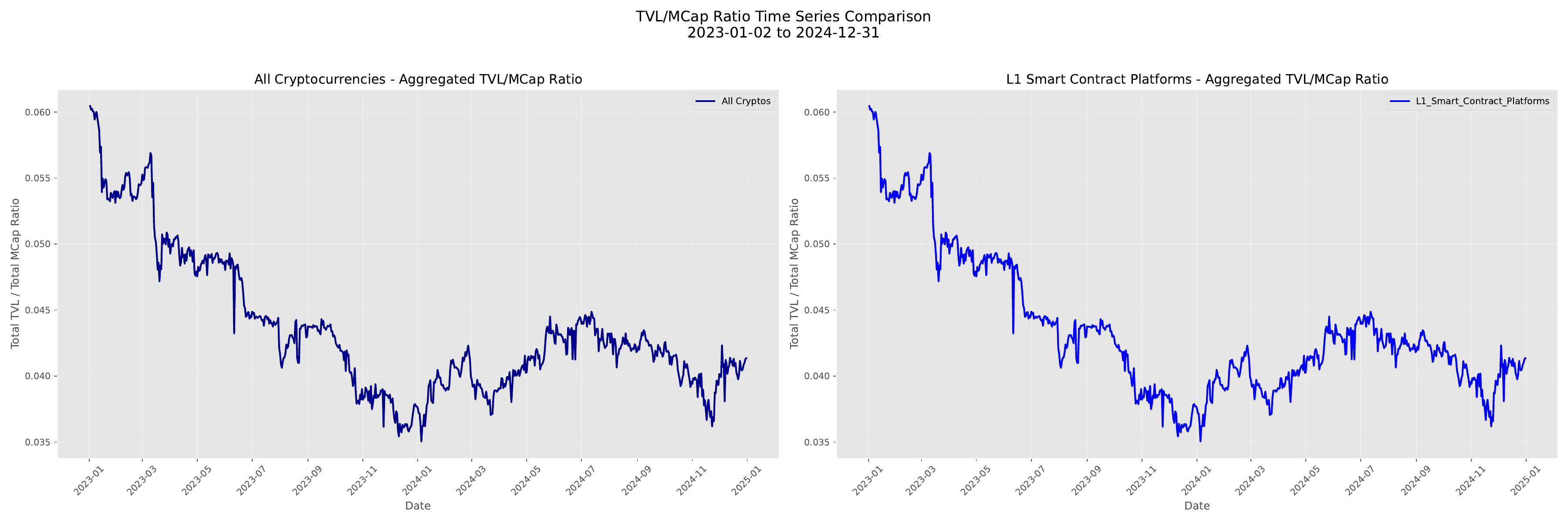}
\caption{\label{fig:org74a7ad0}Simple TVL to Market Capitalization}
\end{figure}
\subsection{Crypto Factor Construction}
\label{sec:org6ed11de}

To construct the TVL factor we first scale TVL by market cap which affords a proportion of market cap which is locked.  Then, for a given week \(t\), we rank assets by TVL / Market Cap in week \(t-1\).  We then create a value-weighted long-short portfolio which buys the top 25\% of coins by TVL and sells the bottom 25\%.  The return on this portfolio over week \(t\) is the factor realization for week \(t\).  This high-minus-low TVL portfolio is denoted as \(HML\) in tables below.  We do this for both total TVL, and a simple TVL measure which excludes values from staking, pool2, governance tokens, borrows, double count, liquid staking, and vesting.  We create this TVL factor for our full set of cryptocurrencies and also our Level 1 only subset.  We also include TVL quartile excess return portfolios. 

To construct the change in TVL to market cap factor we scale TVL by market cap for the previous two weeks.  Then we rank assets by TVL / Market Cap in weeks \(t-1\) and \(t-2\), and calculate the difference of the weeks' TVL to market cap.  This affords the change in TVL to market cap at week \(t-1\).  We then create a value-weighted long-short portfolio which buys the top 25\% of coins by the TVL change and sells the bottom 25\%.  The return on this portfolio over week \(t\) is the factor realization for week \(t\).  This high-minus-low TVL portfolio is denoted as \(HML\) in tables below.  We again do this for our two TVL measures, and our total cryptocurrency and Level 1 only sets.  In addition, we include TVL quartile excess return portfolios. 

To calculate the crypto momentum factor over week \(t\) we calculate the cumulative return for each cryptocurrency over weeks \(t-5\) to \(t-1\).  We then sort the cryptocurrencies and create a value-weighted long-short portfolio which buys the currencies in the top 25\% of cumulative return, and sells the currencies in the bottom 25\%.  The return on this portfolio over week \(t\) is our crypto momentum factor for week \(t\).  We repeat this procedure for all weeks in our sample.

To construct the crypto small-minus-big factor for week \(t\) we sort all cryptocurrencies by size in week \(t-1\).  We then create a value-weighted long-short portfolio which buys the top 25\% of coins by size and sells the bottom 25\%.  The return on this portfolio over week \(t\) is the factor realization for week \(t\).  Lastly, for crypto market returns, we use the weekly total crypto market cap from coinmarketcap.com.  We then subtract the risk-free rate affording excess crypto market returns.
\subsection{Descriptive Statistics}
\label{sec:org18cf0a8}

The goal of the later factor analysis is to regress any portfolio with a statistically significant mean return on our crypto factors. If the intercept (\(\alpha\)) of the regression is also significant, then our model has not explained the anomaly (significant mean excess return).  We therefore first provide descriptive statistics for each portfolio formed on TVL, particularly noting the portfolios with significant mean returns.

Descriptive statistics for TVL HML and quartile portfolios are in the tables below.  Table \ref{tab:org8d82bcc} provides statistics for value-weighted portfolios sorted on total TVL to market capitalization. The TVL HML mean portfolio returns are insignificantly different from 0, however quartiles 1,3, and 4 have positive and significant excess returns.  We therefore test in the next section whether these significant quartile returns can be explained by a crypto factor model, or rather generate alpha.  Plots of annual returns to total TVL to market capitalization by year and functional group are in figures \ref{fig:org8f87995} and \ref{fig:orga4351e9} below.  There is not a clear relationship between the two variables.

\begin{table}[htbp]
\caption{\label{tab:org8d82bcc}All Crypto Value-Weighted Portfolio Descriptive Statistics.  Values are portfolio returns in percentage points and are at the weekly frequency.  Crypto portfolios are formed on total TVL to market cap.  There are 105 weekly observations. \(^{***}\), \(^{**}\), and \(^{*}\), denote statistical significance at the 1\%, 5\%, and 10\% levels respectively.}
\centering
\begin{tabular}{lrrrrr}
\hline
 & HML & Q1 & Q2 & Q3 & Q4\\
\hline
mean & -0.14 & 1.78\(^{**}\) & 1.84 & 1.64\(^{*}\) & 1.64\(^{*}\)\\
std & 5.67 & 7.08 & 11.22 & 9.02 & 8.67\\
min & -14.18 & -14.86 & -22.82 & -24.21 & -18.29\\
25\% & -3.32 & -2.13 & -4.31 & -3.22 & -3.57\\
50\% & -0.71 & 0.04 & -0.49 & 0.51 & 0.49\\
75\% & 1.87 & 5.92 & 6.34 & 7.63 & 6.06\\
max & 19.75 & 26.98 & 44.31 & 31.87 & 28.22\\
\hline
\hline
\end{tabular}
\end{table}

\begin{figure}[htbp]
\centering
\includegraphics[width=.9\linewidth]{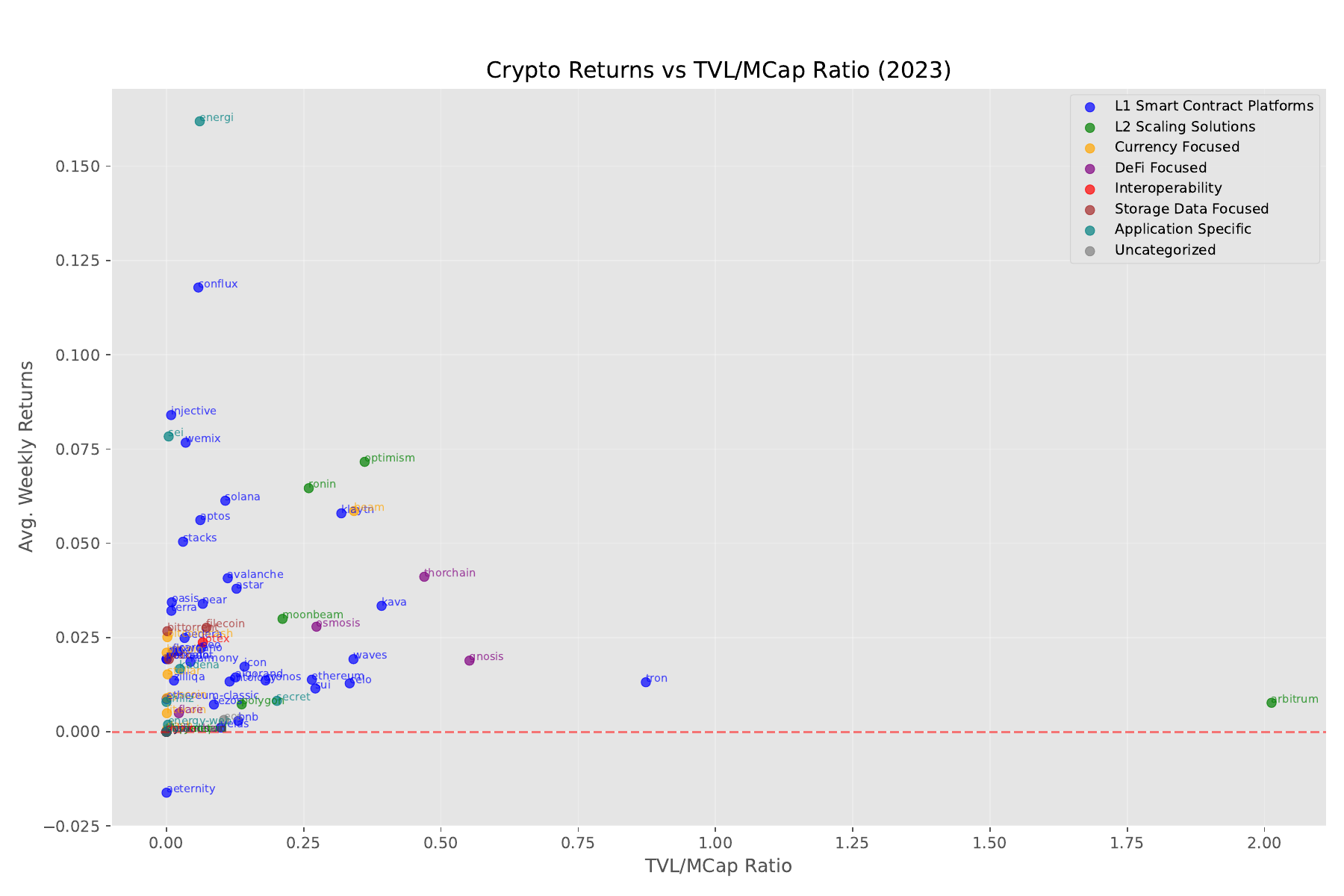}
\caption{\label{fig:org8f87995}Cryptocurrency Returns by Total TVL to Market Capitalization for 2023. Returns are average weekly returns over 2023.}
\end{figure}

\begin{figure}[htbp]
\centering
\includegraphics[width=.9\linewidth]{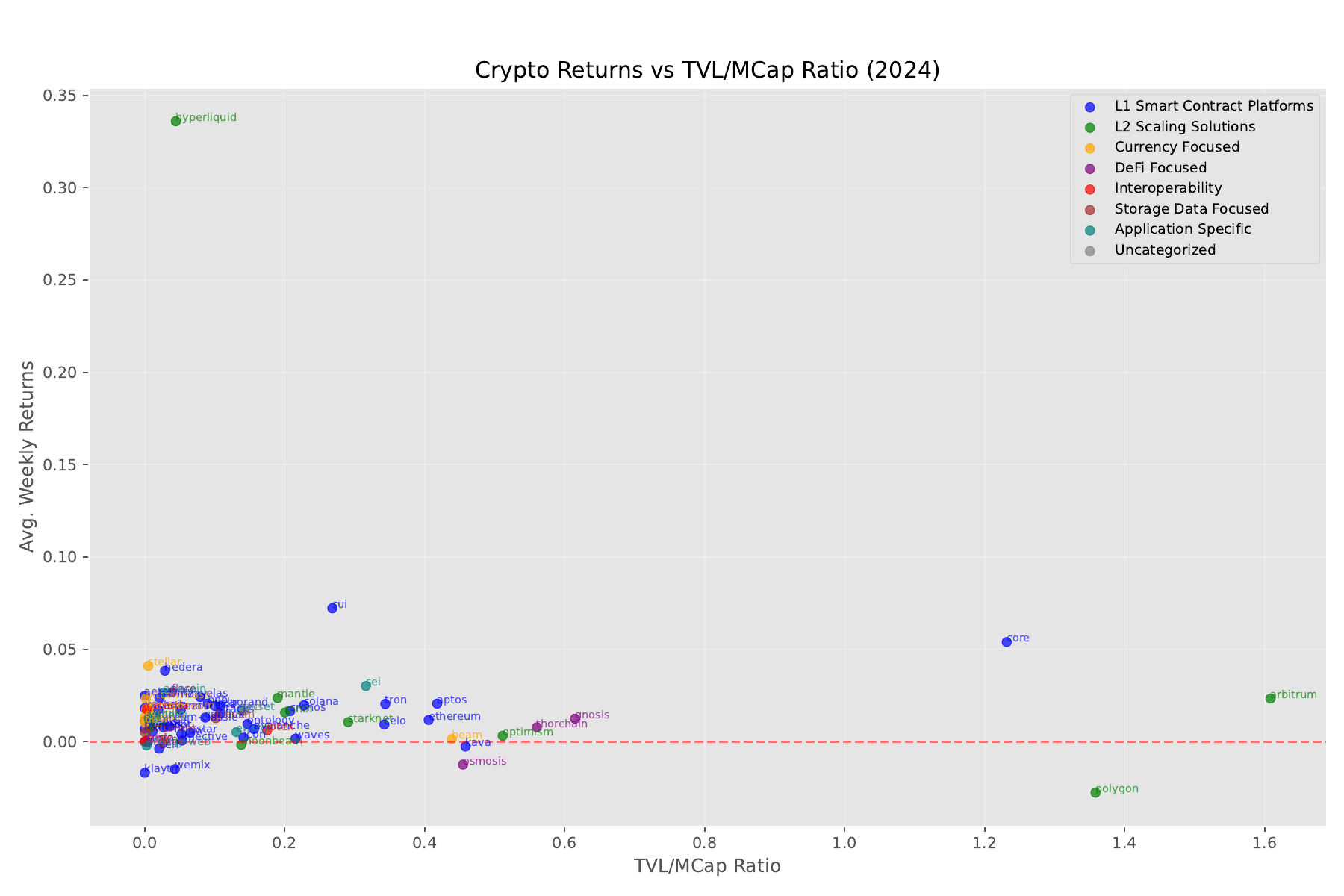}
\caption{\label{fig:orga4351e9}Cryptocurrency Returns by Total TVL to Market Capitalization for 2024. Returns are average weekly returns over 2024.}
\end{figure}

Table \ref{tab:org580f647} shows descriptive statistics for portfolios sorted on the change in TVL to market capitalization. Again the first and third quartile excess return portfolios are significant, however the fourth quartile is not.  The high-minus-low quartile portfolio is insignificant.

\begin{table}[htbp]
\caption{\label{tab:org580f647}All Crypto Value-Weighted Portfolio Descriptive Statistics. Values are portfolio returns in percentage points and are at the weekly frequency.  Crypto portfolios are formed on changes in total TVL to market cap.  There are 104 weekly observations. \(^{***}\), \(^{**}\), and \(^{*}\), denote statistical significance at the 1\%, 5\%, and 10\% levels respectively.}
\centering
\begin{tabular}{lrrrrr}
\hline
 & HML & Q1 & Q2 & Q3 & Q4\\
\hline
mean & -0.51 & 2.04\(^{**}\) & 1.00 & 2.46\(^{**}\) & 1.53\\
std & 7.19 & 9.87 & 10.02 & 10.97 & 9.65\\
min & -26.08 & -27.28 & -18.56 & -21.44 & -23.21\\
25\% & -3.79 & -2.83 & -4.79 & -4.10 & -3.88\\
50\% & -0.31 & 1.38 & -0.45 & 1.18 & -0.12\\
75\% & 3.36 & 7.07 & 6.91 & 6.19 & 6.53\\
max & 17.65 & 36.05 & 43.81 & 52.56 & 35.67\\
\hline
\hline
\end{tabular}
\end{table}

\begin{figure}[htbp]
\centering
\includegraphics[width=.9\linewidth]{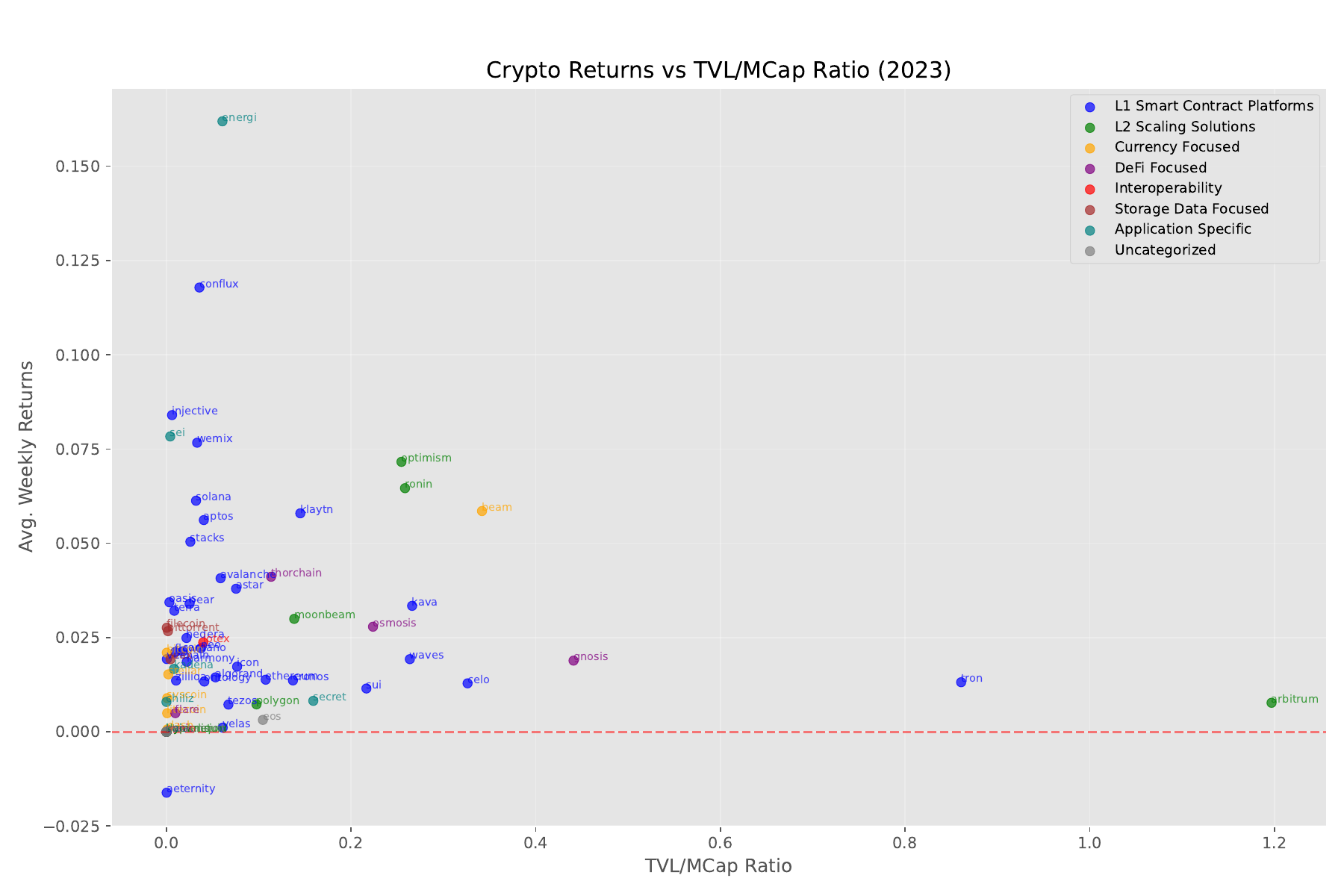}
\caption{\label{fig:orgbec8c7c}Cryptocurrency Returns by Simple TVL to Market Capitalization for 2023. Returns are average weekly returns over 2023.}
\end{figure}

\begin{figure}[htbp]
\centering
\includegraphics[width=.9\linewidth]{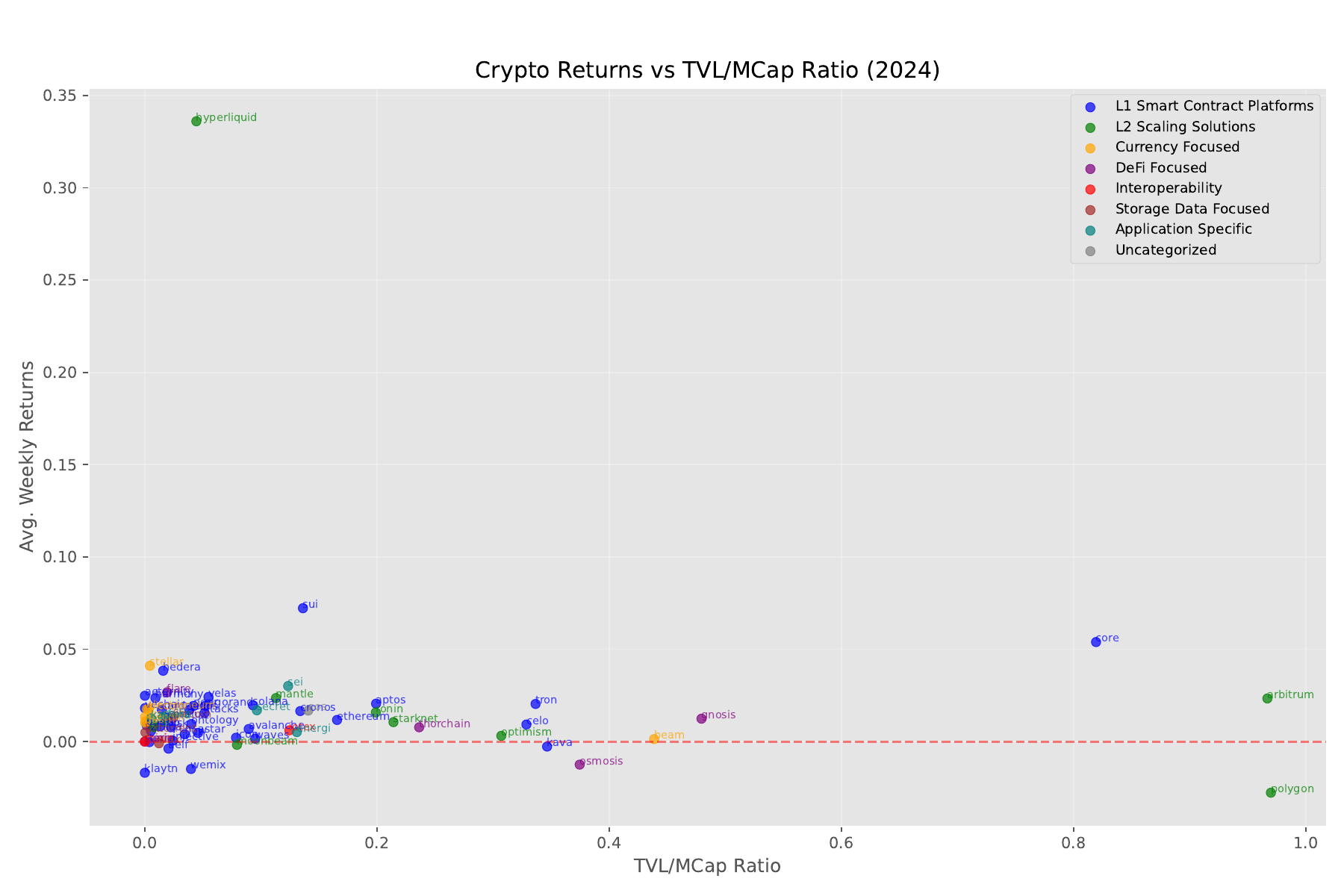}
\caption{\label{fig:org90f83f1}Cryptocurrency Returns by Simple TVL to Market Capitalization for 2024. Returns are average weekly returns over 2024.}
\end{figure}

Table \ref{tab:org9aaa311} shows descriptive statistics for value-weighted TVL HML and quartile excess portfolio returns formed on the simple TVL measure which excludes values from staking, pool2, governance tokens, borrows, double count, liquid staking, and vesting.  Quartiles 2 and 4 returns are significant.  Figures \ref{fig:orgbec8c7c} and \ref{fig:org90f83f1} show annual returns to simple TVL by year.

\begin{table}[htbp]
\caption{\label{tab:org9aaa311}All Crypto Value-Weighted Portfolio Descriptive Statistics.  Values are portfolio returns in percentage points and are at the weekly frequency.  TVL is adjusted for staking, pool2, governance tokens, borrows, double count, liquid staking, and vesting. There are 105 weekly observations. \(^{***}\), \(^{**}\), and \(^{*}\), denote statistical significance at the 1\%, 5\%, and 10\% levels respectively.}
\centering
\begin{tabular}{lrrrrr}
\hline
 & HML & Q1 & Q2 & Q3 & Q4\\
\hline
mean & -0.18 & 1.62 & 2.61\(^{**}\) & 1.19 & 1.44\(^{*}\)\\
std & 7.77 & 11.25 & 13.48 & 8.59 & 7.79\\
min & -61.71 & -22.53 & -27.12 & -19.85 & -17.56\\
25\% & -1.93 & -4.42 & -5.21 & -4.02 & -3.02\\
50\% & 0.63 & -0.04 & 0.72 & 0.12 & 0.88\\
75\% & 3.39 & 6.37 & 10.49 & 6.30 & 6.01\\
max & 13.50 & 71.67 & 51.93 & 30.16 & 25.12\\
\hline
\hline
\end{tabular}
\end{table}

Interestingly, table \ref{tab:org4877923}, which describes portfolios sorted on the change in simple TVL to market capitalization, shows significantly positive mean returns for the cryptocurrencies with the largest, and smallest, changes in TVL.   This indicates that both positive and negative changes in simple TVL lead to positive mean returns.  That is, any large absolute value change in TVL causes positive returns.

\begin{table}[htbp]
\caption{\label{tab:org4877923}All Crypto TVL Change Portfolio Descriptive Statistics.  Values are portfolio returns in percentage points and are at the weekly frequency. TVL is adjusted for staking, pool2, governance tokens, borrows, double count, liquid staking, and vesting. There are 104 weekly observations. \(^{***}\), \(^{**}\), and \(^{*}\), denote statistical significance at the 1\%, 5\%, and 10\% levels respectively.}
\centering
\begin{tabular}{lrrrrr}
\hline
 & HML & Q1 & Q2 & Q3 & Q4\\
\hline
mean & 0.13 & 1.87\(^{*}\) & 1.43 & 1.37 & 2.00\(^{*}\)\\
std & 8.44 & 10.03 & 9.77 & 10.70 & 11.53\\
min & -38.60 & -27.17 & -23.72 & -23.73 & -30.40\\
25\% & -3.07 & -3.35 & -4.23 & -5.27 & -4.95\\
50\% & 0.26 & 0.68 & 0.55 & -0.22 & 0.12\\
75\% & 3.59 & 8.36 & 7.29 & 7.25 & 5.59\\
max & 27.60 & 29.97 & 38.66 & 45.87 & 55.11\\
\hline
\hline
\end{tabular}
\end{table}

\begin{figure}[htbp]
\centering
\includegraphics[width=.9\linewidth]{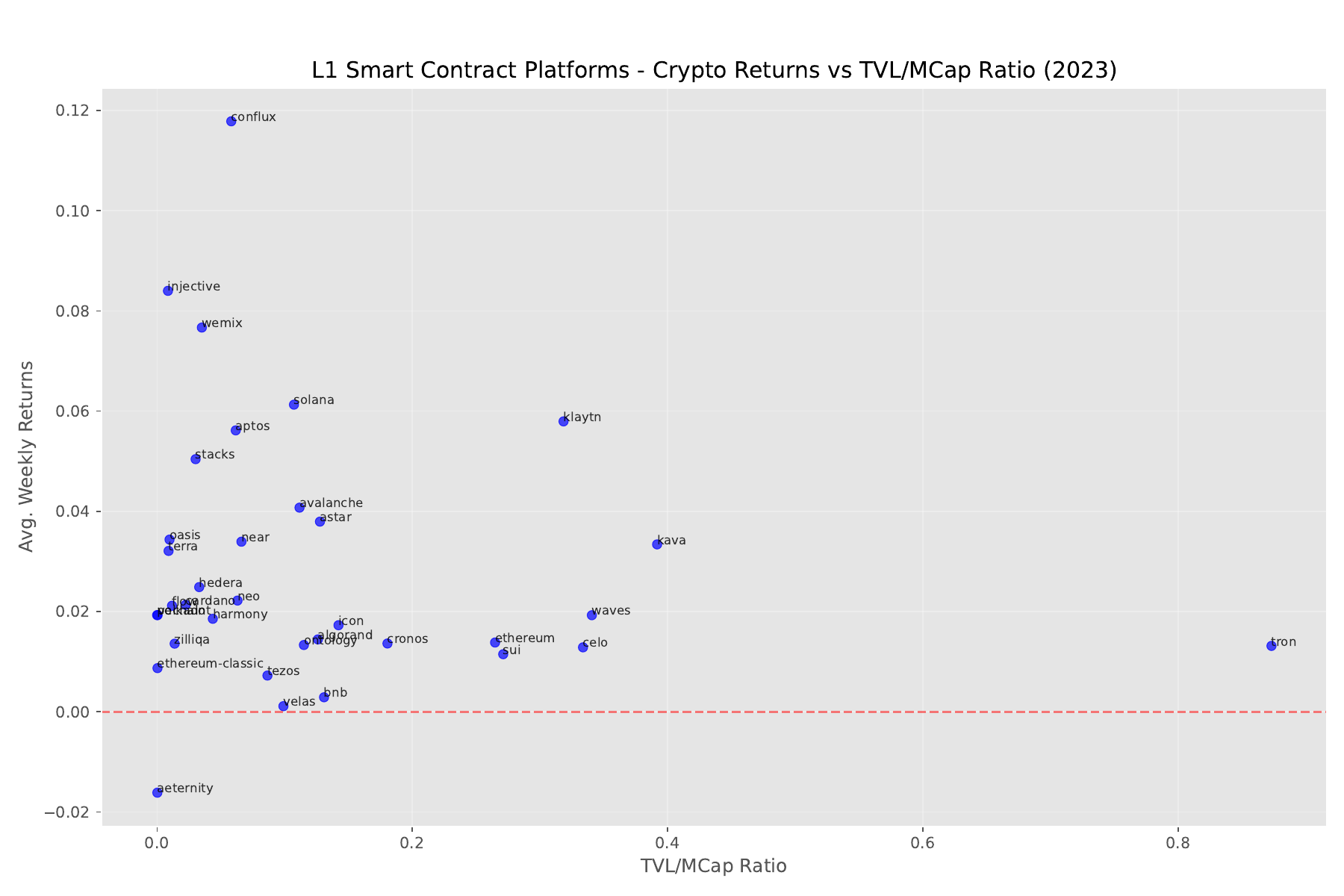}
\caption{\label{fig:org2f5fe6a}Level 1 Cryptocurrency Returns by TVL to Market Capitalization for 2023. Returns are average weekly returns over 2023.}
\end{figure}

\begin{figure}[htbp]
\centering
\includegraphics[width=.9\linewidth]{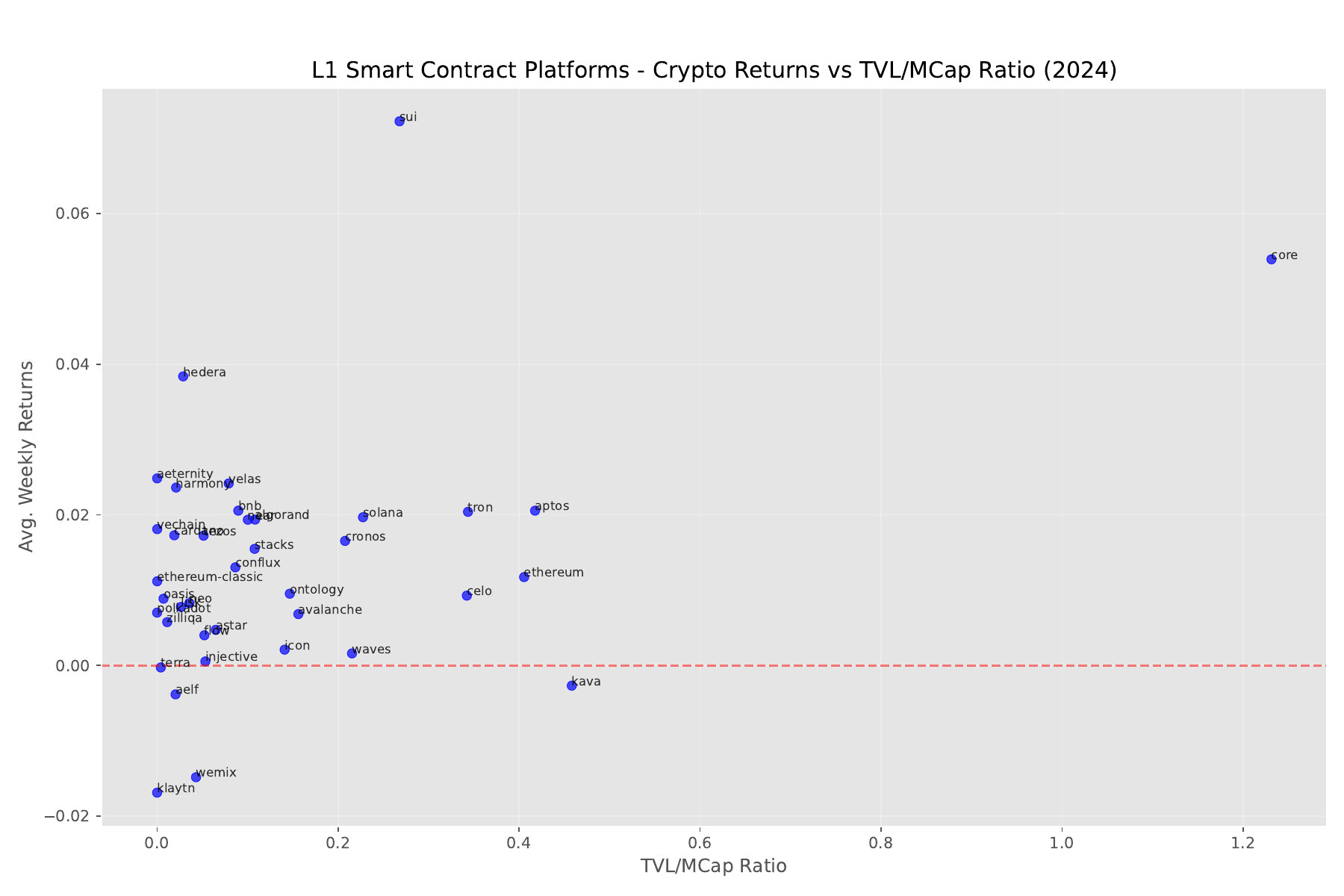}
\caption{\label{fig:org6cda0f4}Level 1 Cryptocurrency Returns by TVL to Market Capitalization for 2024. Returns are average weekly returns over 2024.}
\end{figure}

We also form portfolios based on TVL for a subset of Level 1 tokens.  The descriptive statistics for these portfolios formed on total TVL are in table \ref{tab:org2cddf92} below.  Quartiles 2, 3, and 4 have significantly positive mean excess returns.  However the Level 1 portfolios formed on the change in TVL (table \ref{tab:org73bdf22}) have significant mean returns for quartiles 1, 2, and 4---again it is large changes in absolute value that generate positive returns.  Figures \ref{fig:org2f5fe6a} and \ref{fig:org6cda0f4} show annual returns to total TVL by year for only the level 1 functional group.

\begin{table}[htbp]
\caption{\label{tab:org2cddf92}Level 1 Crypto TVL Portfolio Descriptive Statistics.  Values are portfolio returns in percentage points and are at the weekly frequency.  TVL is total TVL. There are 105 weekly observations. \(^{***}\), \(^{**}\), and \(^{*}\), denote statistical significance at the 1\%, 5\%, and 10\% levels respectively.}
\centering
\begin{tabular}{lrrrrr}
\hline
 & HML & Q1 & Q2 & Q3 & Q4\\
\hline
mean & -0.27 & 1.59 & 2.36\(^{**}\) & 1.70\(^{**}\) & 1.32\(^{*}\)\\
std & 7.79 & 12.19 & 12.49 & 8.99 & 7.90\\
min & -40.92 & -23.27 & -26.04 & -25.07 & -17.73\\
25\% & -2.90 & -5.13 & -4.35 & -2.30 & -3.29\\
50\% & 0.85 & -0.59 & 0.73 & 0.81 & 0.27\\
75\% & 3.63 & 6.62 & 7.60 & 7.77 & 5.98\\
max & 15.72 & 58.49 & 47.24 & 29.04 & 28.99\\
\hline
\hline
\end{tabular}
\end{table}

\begin{table}[htbp]
\caption{\label{tab:org73bdf22}Level 1 Crypto TVL Change Portfolio Descriptive Statistics.  Values are portfolio returns in percentage points and are at the weekly frequency.  TVL is adjusted for staking, pool2, governance tokens, borrows, double count, liquid staking, and vesting. There are 104 weekly observations. \(^{***}\), \(^{**}\), and \(^{*}\), denote statistical significance at the 1\%, 5\%, and 10\% levels respectively.}
\centering
\begin{tabular}{lrrrrr}
\hline
 & HML & Q1 & Q2 & Q3 & Q4\\
\hline
mean & -0.97 & 2.34\(^{*}\) & 2.83\(^{**}\) & 1.59 & 1.37\(^{*}\)\\
std & 8.92 & 13.27 & 13.72 & 12.58 & 8.36\\
min & -36.89 & -26.24 & -28.50 & -47.72 & -17.73\\
25\% & -3.38 & -4.87 & -5.59 & -5.43 & -3.93\\
50\% & 0.25 & -0.38 & 0.21 & 0.84 & 0.29\\
75\% & 3.45 & 8.05 & 10.71 & 8.11 & 6.13\\
max & 22.76 & 65.89 & 48.46 & 56.44 & 28.99\\
\hline
\hline
\end{tabular}
\end{table}

We also provide descriptive statistics for our crypto market factors and the risk-free rate in table \ref{tab:org2dc9e00} below.  The crypto market and momentum portfolios have similar levels of risk and return, however the small-minus-big portfolio has substantially higher risk and lower returns.  This highlights the preference for size often seen in crypto markets.

\begin{table}[htbp]
\caption{\label{tab:org2dc9e00}Crypto Factor and Risk-Free Rate Descriptive Statistics.  Values are returns in percentage points and are at the weekly frequency.  There are 105 weekly observations.}
\centering
\begin{tabular}{lrrrr}
\hline
 & Crypto Market & SMB & Momentum & Risk-Free Rate\\
\hline
mean & 1.44 & -10.04 & 0.68 & 0.09\\
std & 6.39 & 33.46 & 7.68 & 0.01\\
min & -15.42 & -206.53 & -30.14 & 0.06\\
25\% & -1.89 & -6.41 & -4.54 & 0.09\\
50\% & 0.25 & -0.93 & 0.55 & 0.10\\
75\% & 5.72 & 2.18 & 4.33 & 0.10\\
max & 18.70 & 18.54 & 24.93 & 0.11\\
\hline
\hline
\end{tabular}
\end{table}
\section{Method and Results}
\label{sec:orgd40748f}

Our method is similar to that of \cite{FAMA_1996}, which showed that the stock market anomaly portfolios are largely spanned by their three factor model. Their findings suggested that most market anomalies were actually manifestations of the same underlying risk patterns and did not generate abnormal returns.

Similarly, we form portfolios in TVL, and test whether any portfolios with significant mean returns can be spanned by a crypto market model or three factor model.  We test this by estimating the following model, and testing for significant alpha coefficients.  The equation below is the three-factor model, and the market model excludes the \(SMB\) and \(Mom\) factors.

$$r_i - r_f = \alpha_i + \beta_{CM,i} CM + \beta_{SMB,i} SMB + \beta_{Mom,i} Mom + \epsilon_i$$

where \(CM\) denotes value-weighted crypto market returns, \(SMB\) is the crypto market \emph{small-minus-big} portfolio, \(Mom\) is the crypto momentum portfolio, \(r_i\) represents the TVL portfolio returns, and \(r_f\) is the risk-free rate.

We form portfolios on both TVL to market cap, and the change in TVL to market cap.  We also form portfolios using total TVL and a TVL measure which excludes doublecounting and other transactions which overstate TVL.  We find all these portfolios can be spanned by a simple crypto market model.  In a few cases a three-factor crypto factor model improves on the market model, however the improvement is marginal.

Results of regression coefficients for significant TVL portfolio returns on a crypto market model and three factor model are reported below. Alpha coefficients are insignificant across all portfolios formed on TVL and all factor regressions.  Thus we find that all significant mean returns are explained by exposures to common crypto market factors.  In fact, since the alpha coefficients are insignificant across market models, a one factor model is sufficient to explain significant returns of portfolios constructed on TVL.  The crypto market beta is positive and significant in all models. Note, our beta coefficients all tend to be greater than 1 because we exclude Bitcoin from our TVL-formed portfolios, however it dominates the crypto market returns.  For each set of factor models we also use a \cite{Gibbons_1989} test of whether the alpha coefficients are jointly equal to zero (the null hypothesis).  We do not reject the null for any of the \cite{Gibbons_1989} tests, which is evidence that there is no mispricing in our models.

Table \ref{tab:org68ba294} provides factor model results for total TVL to market capitalization.  The models explain from 73\% to 93\% of the variation in the portfolio returns. Interestingly, the beta on the crypto momentum factor is negative and significant for the third quartile portfolio.

\begin{table}[htbp]
\caption{\label{tab:org68ba294}Results from cryptocurrency factor regressions:  \(r_i - r_f = \alpha_i + \beta_{CM,i} CM + \beta_{SMB,i} SMB + \beta_{Mom,i} Mom + \epsilon_i\) where \(i\) denotes the crypto portfolio formed on \(\Delta\frac{TVL}{Market\ Cap}\), \(CM\) is cryptocurrency excess market returns, \(SMB\) is the cryptocurrency small-minus-big factor, \(Mom\) is the cryptocurrency momentum factor, and \(r_f\) is the risk-free rate.  TVL values include all DefiLlama categories.  P-values are below the coefficients in parentheses. \(^{***}\), \(^{**}\), and \(^{*}\), denote statistical significance at the 1\%, 5\%, and 10\% levels respectively.  Test statistics and p-values for the \cite{Gibbons_1989} F-test of whether alpha coefficients are jointly 0 are below the Adjusted-\(R^2\) values.}
\centering
\begin{tabular}{llllllr}
\hline
 & Quartile 1 & Quartile 3 & Quartile 4 & Quartile 1 & Quartile 3 & Quartile 4\\
\hline
\(\alpha\) & 0.25 & -0.11 & -0.04 & 0.31 & -0.05 & -0.04\\
 & (0.20) & (0.80) & (0.93) & (0.13) & (0.92) & (0.93)\\
\(\beta_{CM}\) & 1.06\(^{***}\) & 1.22\(^{***}\) & 1.17\(^{***}\) & 1.07\(^{***}\) & 1.23\(^{***}\) & 1.16\(^{***}\)\\
 & (0.00) & (0.00) & (0.00) & (0.00) & (0.00) & (0.00)\\
\(\beta_{SMB}\) &  &  &  & 0.01 & 0.00 & -0.00\\
 &  &  &  & (0.22) & (0.98) & (0.94)\\
\(\beta_{Mom}\) &  &  &  & 0.02 & -0.11\(^{**}\) & -0.01\\
 &  &  &  & (0.54) & (0.05) & (0.90)\\
\hline
Adj-\(R^2\) & 0.93 & 0.74 & 0.74 & 0.93 & 0.75 & 0.73\\
\hline
 & GRS Stat. & 0.72 &  & GRS Stat. & 1.11 & \\
 & p-value & (0.54) &  & p-value & 0.35 & \\
\hline
\hline
\end{tabular}
\end{table}

Table \ref{tab:org2e06f03} reports results from factor regressions on portfolios formed on the \emph{change} in TVL.  The portfolios with significant mean returns are those portfolios with the most negative, and most positive, changes in TVL to market capitalization.  These portfolios formed on large absolute value changes in TVL are also explained by the factors models (insignificant alpha), however they explain roughly 60\% of the variation in portfolio returns.  There is again evidence on a negative relationship between TVL-formed portfolios and crypto market momentum.

\begin{table}[htbp]
\caption{\label{tab:org2e06f03}Results from cryptocurrency factor regressions:  \(r_i - r_f = \alpha_i + \beta_{CM,i} CM + \beta_{SMB,i} SMB + \beta_{Mom,i} Mom + \epsilon_i\) where \(i\) denotes the crypto portfolio formed on \(\Delta\frac{TVL}{Market\ Cap}\), \(CM\) is cryptocurrency excess market returns, \(SMB\) is the cryptocurrency small-minus-big factor, \(Mom\) is the cryptocurrency momentum factor, and \(r_f\) is the risk-free rate.  TVL includes all optional DeFiLlama categories.  P-values are below the coefficients in parentheses. \(^{***}\), \(^{**}\), and \(^{*}\), denote statistical significance at the 1\%, 5\%, and 10\% levels respectively.  Test statistics and p-values for the \cite{Gibbons_1989} F-test of whether alpha coefficients are jointly 0 are below the Adjusted-\(R^2\) values.}
\centering
\begin{tabular}{lllll}
\hline
 & Quartile 1 & Quartile 4 & Quartile 1 & Quartile 4\\
\hline
\(\alpha\) & 0.23 & 0.38 & 0.41 & 0.21\\
 & (0.70) & (0.54) & (0.50) & (0.65)\\
\(\beta_{CM}\) & 1.23\(^{***}\) & 1.42\(^{***}\) & 1.25\(^{***}\) & 1.41\(^{***}\)\\
 & (0.00) & (0.00) & (0.00) & (0.00)\\
\(\beta_{SMB}\) &  &  & 0.01 & -0.01\\
 &  &  & (0.66) & (0.44)\\
\(\beta_{Mom}\) &  &  & -0.20\(^{***}\) & 0.07\\
 &  &  & (0.01) & (0.39)\\
\hline
Adj-\(R^2\) & 0.64 & 0.68 & 0.66 & 0.68\\
\hline
 & GRS Stat. & 0.23 & GRS Stat. & 0.25\\
 & p-value & (0.80) & p-value & (0.78)\\
\hline
\hline
\end{tabular}
\end{table}

When restricting TVL over all categories (table \ref{tab:org683d28e}), the alpha coefficients for both quartiles with significant mean returns are not significantly different from 0.  The beta coefficients are 1.86 for quartile 2, and 1.01 for quartile 4, and each model explains about 70\% of the variation in the quartile excess returns.  The HML portfolio again had an insignificant mean return.

\begin{table}[htbp]
\caption{\label{tab:org683d28e}Results from cryptocurrency factor regressions:  \(r_i - r_f = \alpha_i + \beta_{CM,i} CM + \beta_{SMB,i} SMB + \beta_{Mom,i} Mom + \epsilon_i\) where \(i\) denotes the crypto portfolio formed on \(\frac{TVL}{Market\ Cap}\), \(CM\) is cryptocurrency excess market returns, \(SMB\) is the cryptocurrency small-minus-big factor, \(Mom\) is the cryptocurrency momentum factor, and \(r_f\) is the risk-free rate. TVL excludes all optional DefiLlama categories.  P-values are below the coefficients in parentheses. \(^{***}\), \(^{**}\), and \(^{*}\), denote statistical significance at the 1\%, 5\%, and 10\% levels respectively.  Test statistics and p-values for the \cite{Gibbons_1989} F-test of whether alpha coefficients are jointly 0 are below the Adjusted-\(R^2\) values.}
\centering
\begin{tabular}{lllll}
\hline
 & Quartile 2 & Quartile 4 & Quartile 2 & Quartile 4\\
\hline
\(\alpha\) & 0.38 & 0.28 & 0.42 & 0.33\\
 & (0.57) & (0.52) & (0.55) & (0.46)\\
\(\beta_{CM}\) & 1.86\(^{***}\) & 1.02\(^{***}\) & 1.86\(^{***}\) & 1.02\(^{***}\)\\
 & (0.00) & (0.00) & (0.00) & (0.00)\\
\(\beta_{SMB}\) &  &  & 0.00 & 0.01\\
 &  &  & (0.89) & (0.63)\\
\(\beta_{Mom}\) &  &  & -0.03 & 0.00\\
 &  &  & (0.73) & (0.95)\\
\hline
Adj-\(R^2\) & 0.76 & 0.70 & 0.76 & 0.69\\
\hline
 & GRS Stat. & 0.27 & GRS Stat. & 0.24\\
 & p-value & (0.77) & p-value & (0.78)\\
\hline
\hline
\end{tabular}
\end{table}

Results of factor model estimation of portfolios formed on the change in simple TVL are similar to the results on the change in total TVL.  Again, the largest TVL changes in absolute value lead to the largest portfolio returns.  Additionally, alpha coefficients are all insignificantly different from zero, and there is evidence of a negative relationship between the crypto momentum portfolio and TVL-sorted portfolio returns.

\begin{table}[htbp]
\caption{\label{tab:org67f6a53}Results from cryptocurrency market model regressions:  \(r_i - r_f = \alpha_i + \beta_{CM,i} CM + \beta_{SMB,i} SMB + \beta_{Mom,i} Mom + \epsilon_i\) where \(i\) denotes the crypto portfolio formed on \(\Delta\frac{TVL}{Market\ Cap}\), \(CM\) is cryptocurrency excess market returns, \(SMB\) is the cryptocurrency small-minus-big factor, \(Mom\) is the cryptocurrency momentum factor, and \(r_f\) is the risk-free rate.  TVL excludes all optional DefiLlama categories.  P-values are below the coefficients in parentheses. \(^{***}\), \(^{**}\), and \(^{*}\), denote statistical significance at the 1\%, 5\%, and 10\% levels respectively.  Test statistics and p-values for the \cite{Gibbons_1989} F-test of whether alpha coefficients are jointly 0 are below the Adjusted-\(R^2\) values.}
\centering
\begin{tabular}{lllll}
\hline
 & Quartile 1 & Quartile 4 & Quartile 1 & Quartile 4\\
\hline
\(\alpha\) & -0.13 & -0.09 & 0.05 & -0.09\\
 & (0.79) & (0.90) & (0.93) & (0.90)\\
\(\beta_{CM}\) & 1.37\(^{***}\) & 1.43\(^{***}\) & 1.38\(^{***}\) & 1.42\(^{***}\)\\
 & (0.00) & (0.00) & (0.00) & (0.00)\\
\(\beta_{SMB}\) &  &  & 0.01 & 0.002\\
 &  &  & (0.42) & (0.92)\\
\(\beta_{Mom}\) &  &  & -0.13\(^{**}\) & 0.04\\
 &  &  & (0.04) & (0.68)\\
\hline
Adj-\(R^2\) & 0.76 & 0.63 & 0.77 & 0.62\\
\hline
 & GRS Stat. & 0.04 & GRS Stat. & 0.01\\
 & p-value & (0.96) & p-value & (0.99)\\
\hline
\hline
\end{tabular}
\end{table}

Results of factor model regressions of the change in total TVL to market cap for level 1 cryptocurrencies are in table \ref{tab:org2cb065a} below.  Results are similar to portfolios formed on the wider set of cryptocurrencies---insignificant alpha coefficients and around 60\% to 80\% of the variation in portfolios returns are explained.

\begin{table}[htbp]
\caption{\label{tab:org2cb065a}Results from cryptocurrency market model regressions on L1 tokens:  \(r_i - r_f = \alpha_i + \beta_i CM + \epsilon_i\) where \(i\) denotes the crypto portfolio, \(CM\) is cryptocurrency excess market returns, and \(r_f\) is the risk-free rate.  TVL values include all DefiLlama categories.  P-values are below the coefficients in parentheses. \(^{***}\), \(^{**}\), and \(^{*}\), denote statistical significance at the 1\%, 5\%, and 10\% levels respectively.  Test statistics and p-values for the \cite{Gibbons_1989} F-test of whether alpha coefficients are jointly 0 are below the Adjusted-\(R^2\) values.}
\centering
\begin{tabular}{llllllr}
\hline
 & Quartile 2 & Quartile 3 & Quartile 4 & Quartile 2 & Quartile 3 & Quartile 4\\
\hline
\(\alpha\) & -0.00 & 0.10 & -0.29 & 0.09 & 0.20 & -0.29\\
 & (0.99) & (0.86) & (0.40) & (0.90) & (0.73) & (0.41)\\
\(\beta_{CM}\) & 1.645\(^{***}\) & 1.11\(^{***}\) & 1.12\(^{***}\) & 1.65\(^{***}\) & 1.12\(^{***}\) & 1.12\(^{***}\)\\
 & (0.00) & (0.00) & (0.00) & (0.00) & (0.00) & (0.00)\\
\(\beta_{SMB}\) &  &  &  & 0.01 & 0.00 & -0.00\\
 &  &  &  & (0.59) & (0.97) & (0.81)\\
\(\beta_{Mom}\) &  &  &  & 0.03 & -0.16\(^{**}\) & -0.03\\
 &  &  &  & (0.77) & (0.02) & (0.48)\\
\hline
Adj-\(R^2\) & 0.71 & 0.62 & 0.82 & 0.70 & 0.63 & 0.81\\
\hline
 & GRS Stat. & 0.25 &  & GRS Stat. & 0.22 & \\
 & p-value & (0.86) &  & p-value & (0.88) & \\
\hline
\hline
\end{tabular}
\end{table}

Results of factor model regressions of the change in simple TVL to market cap for level 1 cryptocurrencies are in table \ref{tab:orgefb048e} below.  The market model is sufficient to explain the significant mean portfolio return, and we see portfolios with smaller changes in TVL have more crypto market risk.

\begin{table}[htbp]
\caption{\label{tab:orgefb048e}Results from cryptocurrency factor model regressions for the Level 1 subset of cryptocurrencies:  \(r_i - r_f = \alpha_i + \beta_{CM,i} CM + \beta_{SMB,i} SMB + \beta_{Mom,i} Mom + \epsilon_i\) where \(i\) denotes the crypto portfolio formed on \(\Delta\frac{TVL}{Market\ Cap}\), \(CM\) is cryptocurrency excess market returns, \(SMB\) is the cryptocurrency small-minus-big factor, \(Mom\) is the cryptocurrency momentum factor, and \(r_f\) is the risk-free rate.  TVL excludes all optional DefiLlama categories.  P-values are below the coefficients in parentheses. \(^{***}\), \(^{**}\), and \(^{*}\), denote statistical significance at the 1\%, 5\%, and 10\% levels respectively.  Test statistics and p-values for the \cite{Gibbons_1989} F-test of whether alpha coefficients are jointly 0 are below the Adjusted-\(R^2\) values.}
\centering
\begin{tabular}{llllllr}
\hline
 & Quartile 1 & Quartile 2 & Quartile 4 & Quartile 1 & Quartile 2 & Quartile 4\\
\hline
\(\alpha\) & -0.09 & 0.21 & -0.31 & -0.14 & 0.24 & -0.30\\
 & (0.91) & (0.78) & (0.45) & (0.87) & (0.75) & (0.48)\\
\(\beta_{CM}\) & 1.66\(^{***}\) & 1.78\(^{***}\) & 1.14\(^{***}\) & 1.65\(^{***}\) & 1.79\(^{***}\) & 1.14\(^{***}\)\\
 & (0.00) & (0.00) & (0.00) & (0.00) & (0.00) & (0.00)\\
\(\beta_{SMB}\) &  &  &  & -0.003 & 0.003 & -0.002\\
 &  &  &  & (0.88) & (0.88) & (0.86)\\
\(\beta_{Mom}\) &  &  &  & 0.03 & -0.01 & -0.04\\
 &  &  &  & (0.77) & (0.96) & (0.40)\\
\hline
Adj-\(R^2\) & 0.64 & 0.69 & 0.77 & 0.63 & 0.69 & 0.76\\
\hline
 & GRS Stat. & 0.24 &  & GRS Stat. & 0.26 & \\
 & p-value & (0.87) &  & p-value & (0.86) & \\
\hline
\hline
\end{tabular}
\end{table}
\section{Conclusion}
\label{sec:org9afcffb}

In this paper we have constructed crypto market and three-factor models, and used them to test the pricing implications of TVL. Our empirical results demonstrate that portfolios formed on the basis of TVL do not exhibit statistically significant returns once adjusted for overall cryptocurrency market performance. Specifically, we find that a single-factor model using the aggregate crypto market return fully explains the cross-section of TVL-sorted portfolio returns.  That is, portfolios created on TVL can be replicated by levered positions in the crypto market portfolio.

These findings suggest that while TVL may proxy for protocol usage or perceived utility, portfolios based on TVL can be successfully priced with standard crypto factor models.  Our results apply to total TVL, and TVL adjusted for staking, pool2, governance tokens, borrows, double count, liquid staking, and vesting.  Future work should consider developing and testing additional measures of locked value, as well as focusing on developing better indicators of user engagement, such as wallet-level activity or unique participants. 

\clearpage

\printbibliography
\end{document}